\documentclass[aps,prd,preprint,superscriptaddress,byrevtex,showpacs,tightenlines]{revtex4}

\usepackage{graphicx}% Include figure files

\begin{document}

% Use the \preprint command to place your local institutional report
% number in the upper righthand corner of the title page in preprint mode.
% Multiple \preprint commands are allowed.
% Use the 'preprintnumbers' class option to override journal defaults
% to display numbers if necessary
\preprint{\vbox{ \hbox{   }
                 \hbox{Belle Preprint 2006-37}
                 \hbox{KEK Preprint 2006-54}
                 \hbox{BELLE-CONF-0668}
%                 \hbox{hep-ex nnnn, if available}
}}

%Title of paper
\title{\quad \\[0.5cm] \boldmath Moments of the Hadronic Invariant
  Mass Spectrum\\
  in $B\to X_c\ell\nu$~Decays at Belle}

% repeat the \author .. \affiliation  etc. as needed
% \email, \thanks, \homepage, \altaffiliation all apply to the current
% author. Explanatory text should go in the []'s, actual e-mail
% address or url should go in the {}'s for \email and \homepage.
% Please use the appropriate macro foreach each type of information

% \affiliation command applies to all authors since the last
% \affiliation command. The \affiliation command should follow the
% other information
% \affiliation can be followed by \email, \homepage, \thanks as well.
%\author{Christoph Schwanda}
%\email[]{Your e-mail address}
%\homepage[]{Your web page}
%\thanks{}
%\altaffiliation{}
%\affiliation{Institute for High Energy Physics, Austrian Academy of
%  Sciences, Vienna}

%Collaboration name if desired (requires use of superscriptaddress
%option in \documentclass). \noaffiliation is required (may also be
%used with the \author command).
%\collaboration can be followed by \email, \homepage, \thanks as well.
\affiliation{Budker Institute of Nuclear Physics, Novosibirsk}
\affiliation{Chiba University, Chiba}
\affiliation{Chonnam National University, Kwangju}
\affiliation{University of Cincinnati, Cincinnati, Ohio 45221}
\affiliation{University of Frankfurt, Frankfurt}
\affiliation{The Graduate University for Advanced Studies, Hayama, Japan} % Sokendai
\affiliation{University of Hawaii, Honolulu, Hawaii 96822}
\affiliation{High Energy Accelerator Research Organization (KEK), Tsukuba}
\affiliation{University of Illinois at Urbana-Champaign, Urbana, Illinois 61801}
\affiliation{Institute of High Energy Physics, Chinese Academy of Sciences, Beijing}
\affiliation{Institute of High Energy Physics, Vienna}
\affiliation{Institute of High Energy Physics, Protvino}
\affiliation{Institute for Theoretical and Experimental Physics, Moscow}
\affiliation{J. Stefan Institute, Ljubljana}
\affiliation{Kanagawa University, Yokohama}
\affiliation{Korea University, Seoul}
\affiliation{Kyungpook National University, Taegu}
\affiliation{Swiss Federal Institute of Technology of Lausanne, EPFL, Lausanne}
\affiliation{University of Ljubljana, Ljubljana}
\affiliation{University of Maribor, Maribor}
\affiliation{University of Melbourne, Victoria}
\affiliation{Nagoya University, Nagoya}
\affiliation{Nara Women's University, Nara}
\affiliation{National Central University, Chung-li}
\affiliation{National United University, Miao Li}
\affiliation{Department of Physics, National Taiwan University, Taipei}
\affiliation{H. Niewodniczanski Institute of Nuclear Physics, Krakow}
\affiliation{Nippon Dental University, Niigata}
\affiliation{Niigata University, Niigata}
\affiliation{University of Nova Gorica, Nova Gorica}
\affiliation{Osaka City University, Osaka}
\affiliation{Osaka University, Osaka}
\affiliation{Panjab University, Chandigarh}
\affiliation{Peking University, Beijing}
\affiliation{RIKEN BNL Research Center, Upton, New York 11973}
\affiliation{University of Science and Technology of China, Hefei}
\affiliation{Seoul National University, Seoul}
\affiliation{Shinshu University, Nagano}
\affiliation{Sungkyunkwan University, Suwon}
\affiliation{University of Sydney, Sydney NSW}
\affiliation{Tata Institute of Fundamental Research, Bombay}
\affiliation{Toho University, Funabashi}
\affiliation{Tohoku Gakuin University, Tagajo}
\affiliation{Tohoku University, Sendai}
\affiliation{Department of Physics, University of Tokyo, Tokyo}
\affiliation{Tokyo Institute of Technology, Tokyo}
\affiliation{Tokyo Metropolitan University, Tokyo}
\affiliation{Virginia Polytechnic Institute and State University, Blacksburg, Virginia 24061}
\affiliation{Yonsei University, Seoul}
  \author{C.~Schwanda}\affiliation{Institute of High Energy Physics, Vienna} % Vienna
  \author{K.~Abe}\affiliation{High Energy Accelerator Research Organization (KEK), Tsukuba} % KEK
  \author{I.~Adachi}\affiliation{High Energy Accelerator Research Organization (KEK), Tsukuba} % KEK
  \author{H.~Aihara}\affiliation{Department of Physics, University of Tokyo, Tokyo} % Tokyo
  \author{D.~Anipko}\affiliation{Budker Institute of Nuclear Physics, Novosibirsk} % BINP
  \author{V.~Aulchenko}\affiliation{Budker Institute of Nuclear Physics, Novosibirsk} % BINP
  \author{E.~Barberio}\affiliation{University of Melbourne, Victoria} % Melbourne
  \author{A.~Bay}\affiliation{Swiss Federal Institute of Technology of Lausanne, EPFL, Lausanne} % Lausanne
  \author{I.~Bedny}\affiliation{Budker Institute of Nuclear Physics, Novosibirsk} % BINP
  \author{K.~Belous}\affiliation{Institute of High Energy Physics, Protvino} % Protvino
% ...
% Ljubljana
\collaboration{The Belle Collaboration}

\begin{abstract}
  We present a measurement of the hadronic invariant mass squared
  ($M^2_X$) spectrum in charmed semileptonic $B$~meson decays $B\to
  X_c\ell\nu$ based on 140 fb$^{-1}$ of Belle data collected near the
  $\Upsilon(4S)$~resonance. We determine the first, the second central
  and the second non-central moments of this spectrum for lepton
  energy thresholds ranging between 0.7 and 1.9 GeV. Full correlations
  between these measurements are evaluated.
\end{abstract}

% insert suggested PACS numbers in braces on next line
\pacs{12.15.Hh,14.40.Nd,13.25.Hw}
% insert suggested keywords - APS authors don't need to do this
%\keywords{}

%\maketitle must follow title, authors, abstract, \pacs, and \keywords
\maketitle

% body of paper here
\section{Introduction}

Inclusive semileptonic decays of $B$~mesons to charmed final states
provide an avenue for measuring the Cabibbo-Kobayashi-Maskawa (CKM)
matrix element~$|V_{cb}|$~\cite{Kobayashi:1973fv} and for determining
non-perturbative hadronic properties of the $B$~meson. In particular,
the moments of the hadronic mass in $B\to X_c\ell\nu$~decays
calculated in the framework of the Operator Product Expansion (OPE)
and the Heavy Quark Effective Theory
(HQET)~\cite{Benson:2003kp,Gambino:2004qm,Benson:2004sg,Bauer:2002sh}
depend on the $b$-quark mass ($m_b$) and a few non-perturbative matrix
elements that also appear in the expression of the total semileptonic
width. Thus, measurements of the hadronic invariant mass
moments~\cite{Aubert:2004te,Csorna:2004kp,Acosta:2005qh,Abdallah:2005cx}
allow the determination of these non-perturbative parameters from the
data and reduce the theoretical uncertainty in the extraction of
$|V_{cb}|$ from measurements of the semileptonic branching fraction. An
improved knowledge of $m_b$ also results in a more precise
determination of $|V_{ub}|$ from inclusive charmless semileptonic
$B$~decays.

This analysis uses $\Upsilon(4S)\to B\bar B$~events in which the
hadronic decay of one $B$~meson is fully reconstructed. The
semileptonic decay of the other $B$ is inferred from the presence of
an identified lepton (electron or muon) amongst the remaining
particles in the event. We calculate the first two moments of the
hadronic invariant mass squared ($M^2_X$) distribution~\cite{ref:0}
directly from the measured spectrum after the effects of finite
detector resolution have been removed using the Singular Value
Decomposition algorithm~\cite{Hocker:1995kb}.

The measurement described in this paper improves the results
previously reported by the BaBar and CLEO
collaborations~\cite{Aubert:2004te,Csorna:2004kp}. The sensitivity to
$m_b$ and other non-perturbative parameters is increased by lowering
the minimum lepton energy threshold to 0.7~GeV. Finally, this analysis
minimizes the dependence on particular $B\to X_c\ell\nu$~model
assumptions by calculating the moments directly from the unfolded
$M^2_X$~spectrum.

\section{Experimental Procedure}

\subsection{Data Sample and Event Selection}

The data used in this analysis were taken with the Belle
detector~\cite{unknown:2000cg} at the KEKB asymmetric energy
$e^+e^-$~collider~\cite{Kurokawa:2001nw}. Belle is a large-solid-angle
magnetic spectrometer that consists of a three-layer silicon vertex
detector, a 50-layer central drift chamber (CDC), an array of
aerogel threshold \v{C}erenkov counters (ACC), a barrel-like
arrangement of time-of-flight scintillation counters (TOF), and an
electromagnetic calorimeter comprised of CsI(Tl) crystals (ECL)
located inside a super-conducting solenoid coil that provides a 1.5~T
magnetic field. An iron flux-return located outside of the coil is
instrumented to detect $K_L^0$ mesons and to identify muons (KLM).

The data sample consists of 140~fb$^{-1}$ taken near the
$\Upsilon(4S)$~resonance, or $152\times 10^6$ $B\bar B$~events. Another
15~fb$^{-1}$ taken at 60~MeV below the resonance are
used to estimate the non-$B\bar B$ (continuum) background. The
off-resonance data is scaled by the integrated on- to off-resonance
luminosity ratio corrected for the $1/s$~dependence of the $q\bar
q$~cross-section.

A generic $B\bar B$~Monte Carlo (MC) sample equivalent to about three
times the integrated luminosity is used in this
analysis. MC-simulated events are generated with
EvtGen~\cite{Lange:2001uf} and full detector simulation based on
GEANT3~\cite{Brun:1987ma} is applied. The decays $B\to D^*\ell\nu$ and
$B\to D\ell\nu$ are generated using an HQET inspired form factor
parameterization~\cite{Caprini:1997mu}. The decays $B\to
D^{**}\ell\nu$~\cite{ref:1} are simulated according to the
Leibovich-Ligeti-Stewart-Wise (LLSW) model~\cite{Leibovich:1997em}
(both relative abundance and form factor shape). The $B\to
X_c\ell\nu$~model also includes non-resonant $B\to
D^{(*)}\pi\ell\nu$~decays which are generated using the Goity-Roberts
model~\cite{Goity:1994xn}. The model for the $B\to
X_u\ell\nu$~background is a hybrid mixture of exclusive modes and an
inclusive component described by the De Fazio-Neubert
model~\cite{DeFazio:1999sv}. Light-cone sum rule form
factors~\cite{Ball:1998kk,Ball:2001fp} are used for $B\to\pi\ell\nu$,
$\rho\ell\nu$ and $\omega\ell\nu$. Other exclusive modes are simulated
according to the ISGW2~model~\cite{Scora:1995ty}. QED bremsstrahlung
in $B\to X\ell\nu$~decays is included using the PHOTOS
package~\cite{Barberio:1993qi}.

Hadronic events are selected based on the charged track multiplicity
and the visible energy in the calorimeter. The selection is described
in detail elsewhere~\cite{Abe:2001hj}.

\subsection{Full-reconstruction Tag}

We fully reconstruct the hadronic decay of one $B$~meson
($B_\mathrm{tag}$) using the decay modes $B^+\to\bar D^{(*)0}\pi^+,
\bar D^{(*)0}\rho^+, \bar D^{(*)0}a_1^+$ and $B^0\to D^{(*)-}\pi^+,
D^{(*)-}\rho^+, D^{(*)-}a_1^+$~\cite{ref:2}. Pairs of photons satisfying
$E_\gamma>50$~MeV in the laboratory-frame and
118~MeV/$c^2<M(\gamma\gamma)<150$~MeV/$c^2$ ($\pm 3.3\sigma$ around
the $\pi^0$~mass) are combined to form
$\pi^0$~candidates. $K^0_S$~mesons are reconstructed from pairs of
oppositely charged tracks with invariant mass within $\pm
30$~MeV/$c^2$ ($\pm 5.1\sigma$) of the nominal $K^0_S$~mass and a
decay vertex displaced from the interaction point. Candidate $\rho^+$
and $\rho^0$~mesons are reconstructed in the $\pi^+\pi^0$ and
$\pi^+\pi^-$ decay modes, requiring their invariant masses to be
within $\pm 150$~MeV/$c^2$ of the nominal $\rho$~mass. Candidate
$a_1^+$~mesons are obtained by combining a $\rho^0$~candidate with a
charged pion and requiring an invariant mass between 1.0 and
1.6~GeV/$c^2$. $D^0$~candidates are searched for in the $K^-\pi^+$,
$K^-\pi^+\pi^0$, $K^-\pi^+\pi^+\pi^-$, $K^0_S\pi^+\pi^-$ and
$K^0_S\pi^0$~decay modes. The $K^-\pi^+\pi^+$ and $K^0_S\pi^+$~modes
are used to reconstruct $D^+$~mesons. Charmed mesons are selected in a
window corresponding to $\pm 3$ times the mass resolution in the
respective decay mode. $D^{*+}$~mesons are reconstructed by pairing a
charmed meson with a low momentum pion, $D^{*+}\to
D^0\pi^+,D^+\pi^0$. The decay modes~$D^{*0}\to D^0\pi^0$ and
$D^{*0}\to D^0\gamma$ are used to search for neutral charmed vector
mesons.

\begin{figure}
  \begin{center}
    \includegraphics{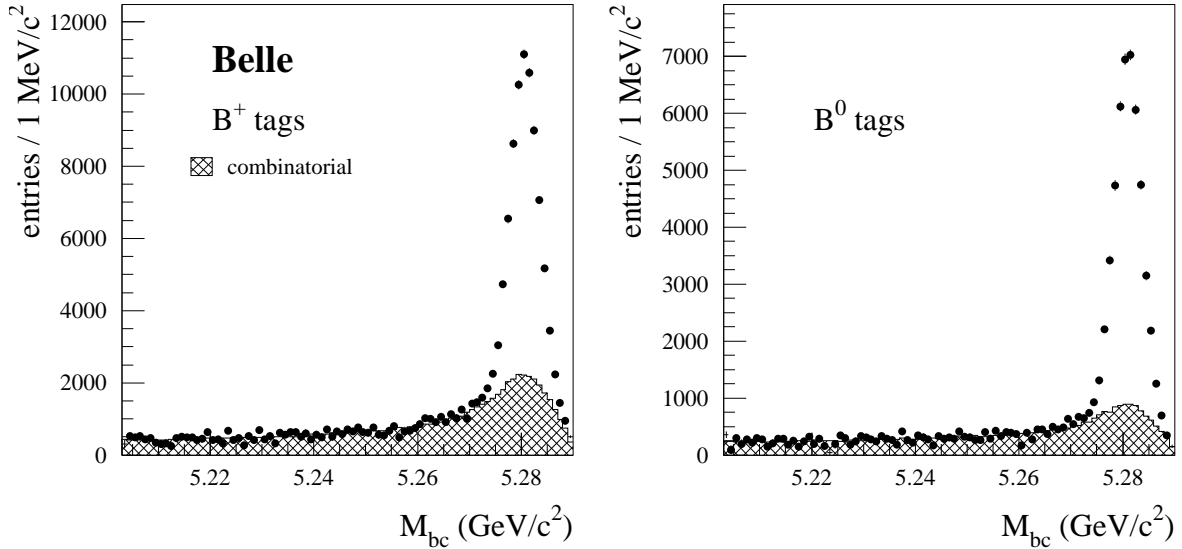}
  \end{center}
  \caption{$M_\mathrm{bc}$~distributions for charged and neutral
    $B_\mathrm{tag}$~candidates after requiring $|\Delta
    E|<0.05$~GeV. No constraints are made on the signal side. The
    points with error bars are on-resonance data after subtraction of
    the scaled off-resonance data. The combinatorial background
    (cross-hatched histogram) is estimated using MC simulation.}
    \label{fig:1}
\end{figure}

For each $B_\mathrm{tag}$~candidate, the beam-energy constrained
mass~$M_\mathrm{bc}$ and the energy difference $\Delta E$ are
calculated,
\begin{equation}
  M_\mathrm{bc} = \sqrt{(E_\mathrm{beam})^2-(\vec p_B)^2}~, \quad
  \Delta E = E_B-E_\mathrm{beam}~,
\end{equation}
where $E_\mathrm{beam}$, $\vec p_B$ and $E_B$ are the beam energy, the
3-momentum and the energy of the $B$~candidate in the
$\Upsilon(4S)$~frame. In $M_\mathrm{bc}$ and $\Delta E$, the signal
peaks at the nominal $B$~mass and zero, respectively. We define the
signal region by the selections $M_\mathrm{bc}>5.27$~GeV/$c^2$ and
$|\Delta E|<0.05$~GeV. If multiple candidates are found in a single
event, the best candidate is chosen based on the proximity of $\Delta
E$, $M_D$ and $\Delta M$ to their nominal values, where $M_D$ is the
reconstructed $D$~meson mass and $\Delta M$ is the difference between
the reconstructed $D^*$ and $D$~meson masses. Without making any
requirement on the decay of the other $B$~meson, the number of $B^+$
($B^0$) tags in this region, after subtraction of continuum and
combinatorial backgrounds, is $61,365\pm 531$ ($41,027\pm 368$),
Fig.~\ref{fig:1}.

\subsection{Lepton Reconstruction}

Semileptonic decays of the other $B$~meson ($B_\mathrm{signal}$) are
selected by searching for an identified charged lepton (electron or
muon) within the remaining particles in the event. Electron candidates
are identified using the ratio of the energy detected in the ECL to
the track momentum, the ECL shower shape, position matching between
track and ECL cluster, the energy loss in the CDC and the response of
the ACC~counters. Muons are identified based on their penetration
range and transverse scattering in the KLM~detector. In the momentum
region relevant to this analysis, charged leptons are identified with
an efficiency of about 90\% and the probability to misidentify a pion
as an electron (muon) is 0.25\%
(1.4\%)~\cite{Hanagaki:2001fz,Abashian:2002bd}.

We further require electron (muon) candidates to originate from near
the interaction vertex, have a laboratory-frame momentum greater than
0.3~GeV/$c$ (0.6~GeV/$c$) and satisfy $17^\circ<\theta<150^\circ$
($25^\circ<\theta<145^\circ$), where $\theta$ is the polar angle in
the laboratory-frame relative to the beam direction. If more than one
charged lepton candidate is found in the event, we only keep the one
with the highest momentum in the $B$~rest frame. Electrons from photon
conversion are vetoed by rejecting the event if the invariant mass of
the electron candidate and another oppositely charged particle in the
event is below 0.04~GeV/$c^2$ and secondary vertex criteria are
satisfied. If the charged lepton candidate is consistent with the decay
$J/\psi\to\ell^+\ell^-$ ({\it i.e.}, the invariant mass of the lepton
candidate and another oppositely charged lepton in the event is
between 3~GeV/$c^2$ and 3.15~GeV/$c^2$), the event is also rejected.

In $B^+$~tagged events, we require the lepton charge to be consistent
with a prompt semileptonic decay of $B_\mathrm{signal}$. In
$B^0$~events, we make no requirement on the lepton charge. In electron
events, we partially recover the effect of bremsstrahlung by searching
for a photon with laboratory-frame energy $E_\gamma<1$~GeV within a
5$^\circ$~cone around the electron direction at the interaction
point. If such a photon is found, it is merged with the electron and
removed from the event.

\subsection{Hadronic Mass Reconstruction}

The 4-momentum $p_X$ of the hadronic system~$X$ recoiling against
$\ell\nu$ is determined by summing the 4-momenta of the remaining
charged tracks and unmatched clusters in the event. We exclude tracks
passing very far away from the interaction point or compatible with a
multiply reconstructed track generated by a low-momentum particle
spiraling in the central drift chamber. Unmatched clusters in the
barrel region must have an energy greater than 50~MeV. Higher
thresholds are applied in the endcap regions.

To improve the resolution in $M^2_X$, we reject events with a missing
mass larger than 3~GeV$^2$/$c^4$. Further improvement is obtained by
recalculating the 4-momentum of the $X$~system,
\begin{equation}
  p'_X =
  (p_{e^+\mathrm{-beam}}+p_{e^-\mathrm{-beam}})-p_{B_{tag}}-p_\ell-p_\nu~,
\end{equation}
taking the neutrino 4-momentum $(E_\nu,\vec p_\nu)$ to be $(|\vec
p_\mathrm{miss}|,\vec p_\mathrm{miss})$, where $\vec p_\mathrm{miss}$
is the missing 3-momentum. Defined as the half width at half maximum,
the resolution in $M^2_X$ obtained from $p'_X$ is about
0.8~GeV$^2/c^4$, compared to 1.4~GeV$^2/c^4$ in $M^2_X$ from $p_X$.

\subsection{Backgrounds in the Hadronic Mass Spectrum}

We consider the following contributions to the background in the
$M^2_X$~spectrum: non-$B\bar B$ (continuum) background, combinatorial
background, background from secondary or fake leptons and $B\to
X_u\ell\nu$~background. Combinatorial background are true $B\bar
B$~events for which reconstruction or flavor assignment of the tagged
$B$~meson is not correct.

The shapes of these background components in $M^2_X$ are determined
from the MC simulation, except for the continuum where off-resonance
data is used. The shape of the fake muon background is corrected by
the ratio of the pion fake rate in the experimental data over the same
quantity in the MC simulation, as measured using kinematically
identified pions in $K^0_S\to\pi^+\pi^-$~decays. We derive the shape
of the combinatorial background from the generic $B\bar B$~simulation
by selecting events in which the reconstruction of $B_\mathrm{tag}$ does
not correspond precisely to what was generated in the simulation.

The continuum background is scaled by the integrated on- to
off-resonance luminosity ratio, taking into account the cross-section
difference. The MC-prediction of the combinatorial background is
normalized to the data using the side-band region
($M_\mathrm{bc}>5.27$~GeV/$c^2$ and $0.15<|\Delta E|<0.3$~GeV). The
normalization of the secondary or fake lepton background is found from
the data by fitting the electron (muon) momentum
distribution~$p^*_\ell$~\cite{ref:3} in the $B$~meson rest frame in
the range from 0.3 to 2.4~GeV/$c$ (0.6 to 2.4~GeV/$c$). The
$X_u\ell\nu$~component is normalized to the number of $B^+$ ($B^0$)
tags, assuming a branching fraction of $2.08\times 10^{-3}$
($1.92\times 10^{-3}$) for $B^+\to X^0_u\ell^+\nu$ ($B^0\to
X^-_u\ell^+\nu$)~\cite{ref:4}.

The background in the $M^2_X$~spectrum is estimated separately in the
four sub-samples, defined by the charge of $B_\mathrm{tag}$ ($B^+$,
$B^0$) and the lepton type (electron, muon).

The purity of the $B\to X_c\ell\nu$~signal depends on the sub-sample
and the lepton energy threshold, typical values being around
75\%. Table~\ref{tab:1} shows the numbers of signal events and
purities for each combination of $B_\mathrm{tag}$~charge, lepton type
and lepton energy threshold.
\begin{table}
  \caption{Number of $B\to X_c\ell\nu$~signal candidates and signal
    purity in the
    four sub-samples, as a function of the lepton energy
    threshold. The yields are quoted with their statistical
    uncertainty; the corresponding signal purity is given in
    parentheses.} \label{tab:1}
  \begin{center}
    \begin{tabular}{c@{\extracolsep{.3cm}}cccc}
      \hline \hline
      \rule[-1.3ex]{0pt}{4ex}$E^*_\mathrm{min}$ & $B^+$ electron &
      $B^+$ muon & $B^0$ electron & $B^0$ muon\\
      \hline
      \rule{0pt}{2.7ex}0.7 & $4105\pm 100$ (70.5\%) & $3739\pm 108$
      (61.5\%) & $2491\pm 80$ (65.9\%) & $2400\pm 86$ (60.3\%)\\
      0.9 & $3855\pm \phantom{1}95$ (73.2\%) & $3591\pm 104$ (64.8\%)
      & $2353\pm 76$ (73.4\%) & $2307\pm 83$ (67.3\%)\\
      1.1 & $3466\pm \phantom{1}86$ (74.9\%) & $3305\pm \phantom{1}96$
      (68.3\%) & $2098\pm 68$ (77.1\%) & $2120\pm 76$ (74.2\%)\\
      1.3 & $2894\pm \phantom{1}72$ (75.8\%) & $2857\pm \phantom{1}84$
      (70.6\%) & $1749\pm 58$ (80.4\%) & $1800\pm 66$ (78.0\%)\\
      1.5 & $2195\pm \phantom{1}56$ (74.6\%) & $2225\pm \phantom{1}66$
      (72.3\%) & $1322\pm 45$ (84.2\%) & $1388\pm 52$ (79.7\%)\\
      1.7 & $1384\pm \phantom{1}38$ (77.2\%) & $1415\pm \phantom{1}44$
      (72.4\%) & $\phantom{1}824\pm 30$ (83.7\%) & $\phantom{1}878\pm
      34$ (80.7\%)\\
      \rule[-1.3ex]{0pt}{1.3ex}1.9 & $\phantom{1}571\pm \phantom{1}19$
      (73.8\%) & $\phantom{1}627\pm \phantom{1}22$ (74.0\%) &
      $\phantom{1}353\pm 15$ (84.3\%) & $\phantom{1}376\pm 17$
      (76.7\%)\\
      \hline \hline
    \end{tabular}
  \end{center}
\end{table}

\subsection{Unfolding and Moment Calculation}

\begin{figure}
  \begin{center}
    \includegraphics{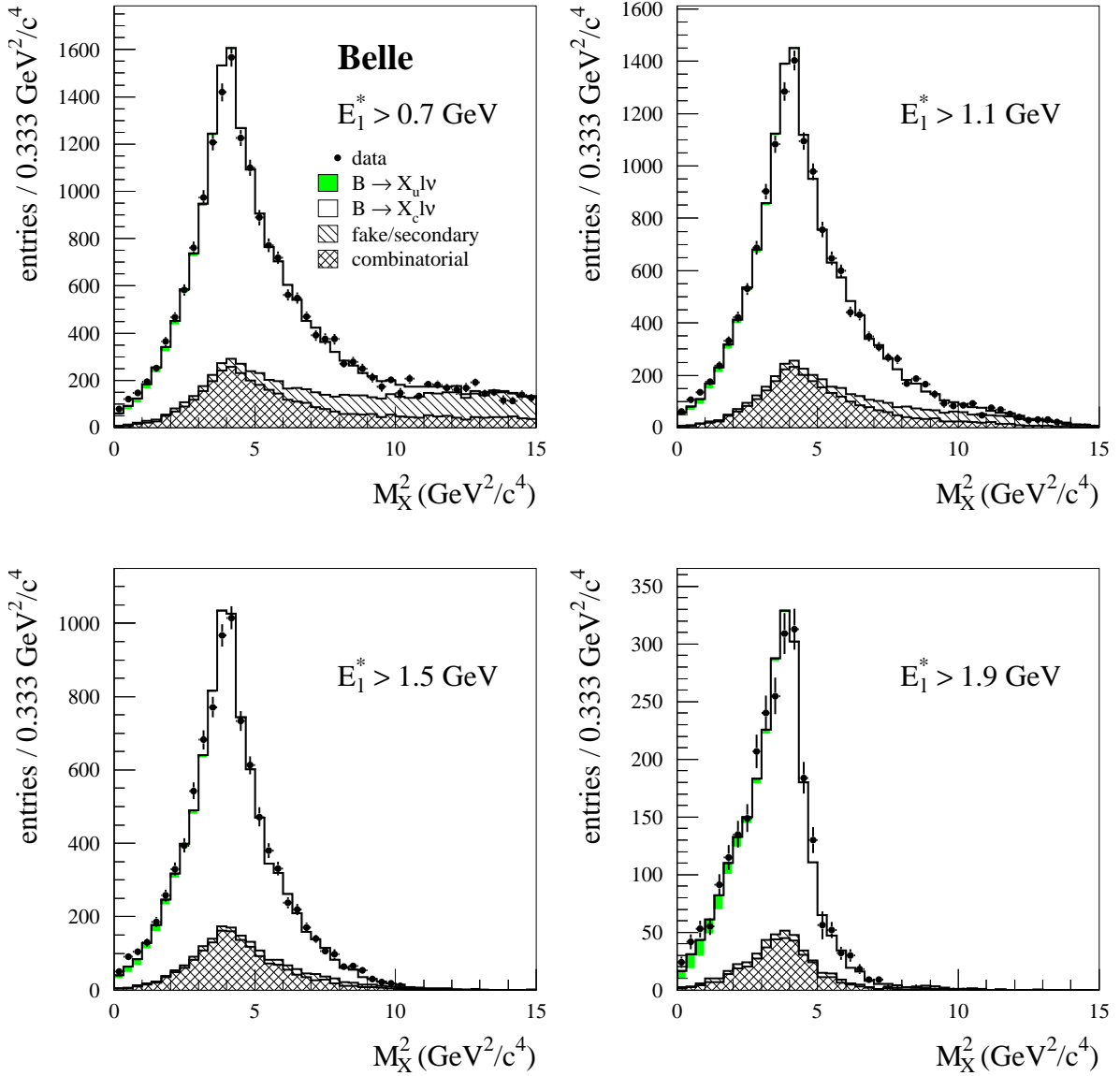}
  \end{center}
  \caption{Measured hadronic mass spectrum for different lepton energy
  thresholds. The points with error bars are the experimental data
  after subtraction of the continuum background. The histograms show the $B\to
  X_c\ell\nu$~signal and the different background components,
  explained in more detail in the text.} \label{fig:2}
\end{figure}

We measure the $M^2_X$~spectrum in 45~bins in the range from 0 to
15~GeV$^2$/$c^4$ (bin width 0.333 GeV$^2$/$c^4$), which is shown in
Fig.~\ref{fig:2}, and unfold the finite detector resolution in this
distribution using the Singular Value Decomposition (SVD)
algorithm~\cite{Hocker:1995kb}. The
unfolded $M^2_X$~spectrum has 15~bins in the range from $M^2_D$ to
about 15 GeV$^2/c^4$. The bin width is 1~GeV$^2/c^4$, except around
the narrow states -- $D$, $D^*$, $D_1$ and $D^*_2$ -- where smaller
bin sizes are chosen.

The unfolding is done separately in each sub-sample ($B^+$ electron,
$B^+$ muon, $B^0$ electron and $B^0$ muon). From the unfolded
spectrum, we calculate the first moment and its statistical
uncertainty squared,
\begin{equation}
  \langle M^2_X\rangle = \frac{\sum_i(M^2_X)_ix'_i}{\sum_ix'_i}~, \quad
  \sigma^2(\langle M^2_X\rangle) =
  \frac{\sum_{i,j}(M^2_X)_iX_{ij}(M^2_X)_j}{(\sum_ix'_i)^2}~. \label{eq:1}
\end{equation}
Here, $x'$ is the unfolded spectrum corrected for slightly different
bin-to-bin efficiencies and $X$ is its covariance matrix, also
determined by the SVD algorithm. $(M^2_X)_i$ is the central value of
the $i$-th bin of the unfolded spectrum. The second central and
non-central moments, $\langle (M^2_X-\langle M^2_X\rangle)^2\rangle$
and $\langle M^4_X\rangle$ are calculated from the same spectrum,
substituting $M^2_X$ by $(M^2_X-\langle M^2_X\rangle)^2$ and $M^4_X$
in Eq.~\ref{eq:1}, respectively.

As the hadron mass moments are not expected to depend on the $B$~meson
charge or the lepton type~\cite{Gambino:2004qm,Bauer:2002sh}, we take
the average over the four sub-sample results.

We have tested the entire measurement procedure including event
reconstruction, unfolding and moment calculation on MC simulated
events and no significant bias has been observed over the full range
of lepton energy thresholds.

\section{Results and Systematic uncertainties}

\subsection{Results}

\begin{figure}
  \begin{center}
    \includegraphics{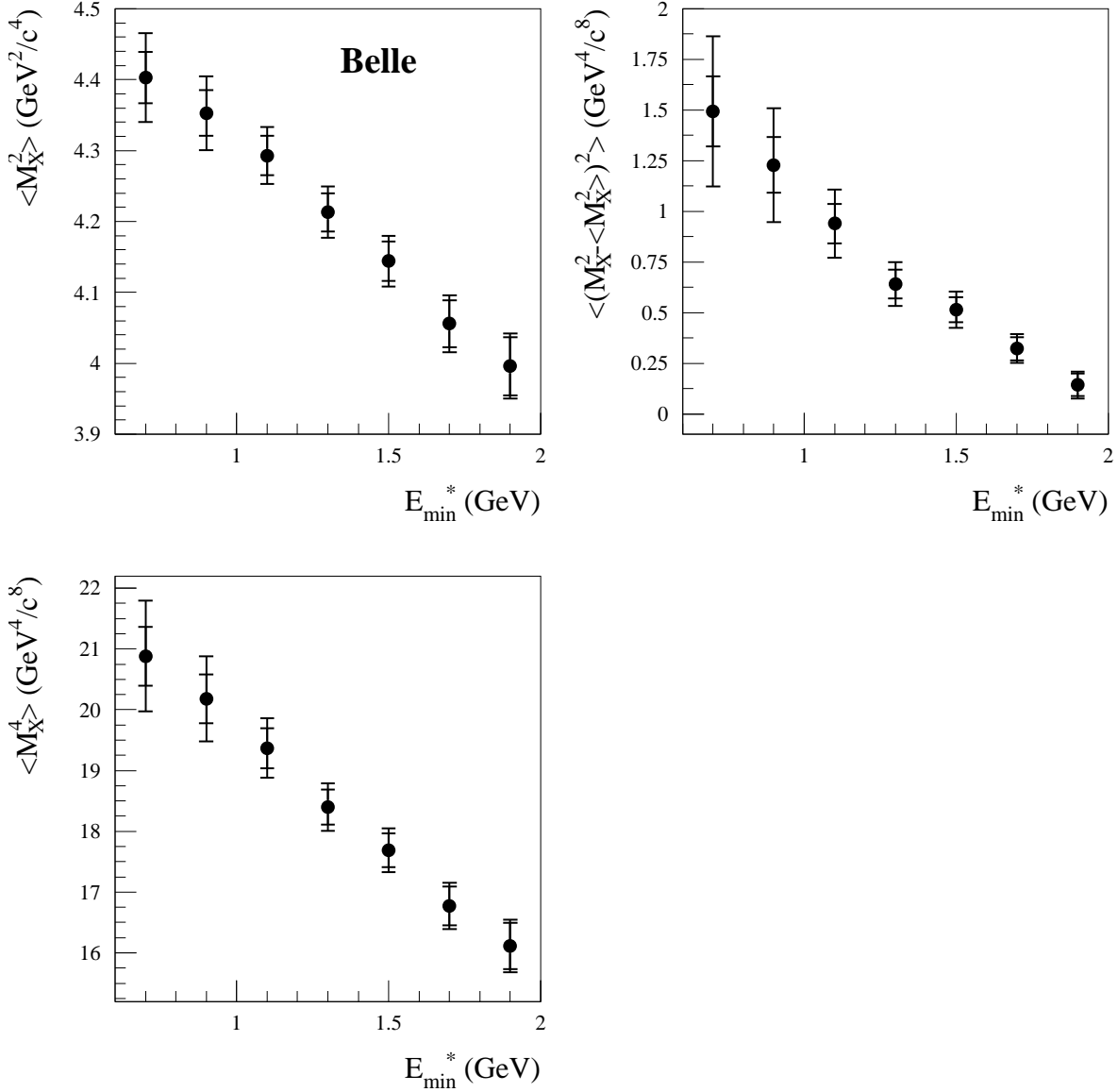}
  \end{center}
  \caption{Graphical representation of the results in
   Table~\ref{tab:2}. The error bars indicate the statistical and
   total experimental errors.} \label{fig:3}
\end{figure}

Our measurements of $\langle M^2_X\rangle$, $\langle(M^2_X-\langle
M^2_X\rangle)^2\rangle$ and $\langle M^4_X\rangle$ for different
lepton energy thresholds are shown in Table~\ref{tab:2} and
Fig.~\ref{fig:3}. The sub-sample results for a given charge of
$B_\mathrm{tag}$ ($B^+$, $B^0$) or lepton type (electron, muon) are
compatible within their statistical uncertainty only.
\begin{table}
  \caption{Measurements of $\langle M^2_X\rangle$,
    $\langle(M^2_X-\langle M^2_X\rangle)^2\rangle$ and $\langle
    M^4_X\rangle$ for different lepton energy thresholds. The results
    in this table are the averages of the four sub-samples, defined by
    the charge of $B_\mathrm{tag}$ ($B^+$, $B^0$) and the lepton type
    (electron, muon). The first error is statistical, the second is
    the estimated systematic uncertainty. The different measurements
    are highly correlated (Tables~\ref{tab:6}--\ref{tab:10}).}
  \label{tab:2}
  \begin{center}
    \begin{tabular}{c@{\extracolsep{.3cm}}ccc}
      \hline \hline
      \rule[-1.3ex]{0pt}{4ex}$E^*_\mathrm{min}$ (GeV) & $\langle
      M^2_X\rangle$ (GeV$^2$/$c^4$) & $\langle(M^2_X-\langle
      M^2_X\rangle)^2\rangle$ (GeV$^4$/$c^8$) & $\langle M^4_X\rangle$
      (GeV$^4$/$c^8$)\\
      \hline
      \rule{0pt}{2.7ex}0.7 & $4.403\pm 0.036\pm 0.052$ & $1.494\pm
      0.173\pm 0.327$ & $20.88\pm  0.48\pm 0.77$\\
      0.9 & $4.353\pm 0.032\pm 0.041$ & $1.229\pm 0.138\pm 0.244$ &
      $20.18\pm  0.40\pm 0.58$\\
      1.1 & $4.293\pm 0.028\pm 0.029$ & $0.940\pm 0.098\pm 0.137$ &
      $19.37\pm  0.33\pm 0.36$\\
      1.3 & $4.213\pm 0.027\pm 0.024$ & $0.641\pm 0.071\pm 0.080$ &
      $18.40\pm  0.29\pm 0.26$\\
      1.5 & $4.144\pm 0.028\pm 0.022$ & $0.515\pm 0.061\pm 0.064$ &
      $17.69\pm  0.28\pm 0.23$\\
      1.7 & $4.056\pm 0.033\pm 0.022$ & $0.322\pm 0.058\pm 0.040$ &
      $16.77\pm  0.32\pm 0.21$\\
      \rule[-1.3ex]{0pt}{1.3ex}1.9 & $3.996\pm 0.041\pm 0.021$ &
      $0.143\pm 0.056\pm 0.038$ & $16.11\pm  0.38\pm 0.20$\\
      \hline \hline
    \end{tabular}
  \end{center}
\end{table}

\subsection{Systematic Uncertainties}

The different contributions to the systematic error are shown in
Tables~\ref{tab:3}--\ref{tab:5}. The total systematic error in
Table~\ref{tab:2} corresponds to the quadratic sum of these
components.
\begin{table}
  \caption{Breakup of the systematic error on $\langle
    M^2_X\rangle$. Refer to the text for details.} \label{tab:3}
  \begin{center}
    \begin{tabular}{l@{\extracolsep{.3cm}}ccccccc}
      \hline \hline
      \rule{0pt}{2.7ex} & \multicolumn{7}{c}{$\Delta\langle
      M^2_X\rangle$ (GeV$^2$/$c^4$)}\\
      \rule[-1.3ex]{0pt}{1.3ex}$E^*_\mathrm{min}$ (GeV) & 0.7 & 0.9 &
      1.1 & 1.3 & 1.5 & 1.7 & 1.9\\
      \hline
      \rule{0pt}{2.7ex}secondary/fake leptons & 0.033 & 0.023 & 0.013
      & 0.007 & 0.004 & 0.002 & 0.000\\
      combinatorial background & 0.006 & 0.004 & 0.003 & 0.002 & 0.002
      & 0.002 & 0.000\\
      continuum & 0.000 & 0.000 & 0.000 & 0.000 & 0.000 & 0.000 &
      0.000\\
      \rule[-1.3ex]{0pt}{1.3ex}$B\to X_u\ell\nu$ background & 0.004 &
      0.004 & 0.004 & 0.004 & 0.006 & 0.007 & 0.009\\
      \hline
      \rule{0pt}{2.7ex}$\mathcal{B}(D^{(*)}\ell\nu)$ & 0.008 & 0.007 &
      0.007 & 0.007 & 0.006 & 0.005 & 0.003\\
      $\mathcal{B}(D^{**}\ell\nu)$ & 0.022 & 0.014 & 0.006 & 0.000 &
      0.000 & 0.008 & 0.006\\
      $\mathcal{B}((D^{(*)}\pi)_\mathrm{non-res.}\ell\nu)$ & 0.024 &
      0.017 & 0.007 & 0.004 & 0.004 & 0.004 & 0.004\\
      $D^{(*)}\ell\nu$ form factors & 0.013 & 0.013 & 0.012 & 0.011 &
      0.010 & 0.008 & 0.006\\
      \rule[-1.3ex]{0pt}{1.3ex}$D^{**}\ell\nu$ form factors & 0.003 &
      0.002 & 0.002 & 0.001 & 0.001 & 0.001 & 0.004\\
      \hline
      \rule{0pt}{2.7ex}unfolding & 0.015 & 0.015 & 0.015 & 0.015 &
      0.015 & 0.015 & 0.015\\
      binning & 0.001 & 0.001 & 0.001 & 0.001 & 0.001 & 0.000 &
      0.001\\
      \rule[-1.3ex]{0pt}{1.3ex}efficiency & 0.008 & 0.011 & 0.012 &
      0.009 & 0.008 & 0.005 & 0.004\\
      \hline
      \rule[-1.3ex]{0pt}{4ex}total & 0.052 & 0.041 & 0.029 & 0.024 &
      0.022 & 0.022 & 0.021\\
      \hline \hline
    \end{tabular}
  \end{center}
\end{table}
\begin{table}
  \caption{Same as Table~\ref{tab:3} for $\langle(M^2_X-\langle
    M^2_X\rangle)^2\rangle$.}
  \begin{center}
    \begin{tabular}{l@{\extracolsep{.3cm}}ccccccc}
      \hline \hline
      \rule{0pt}{2.7ex} &
      \multicolumn{7}{c}{$\Delta\langle(M^2_X-\langle
	M^2_X\rangle)^2\rangle$ (GeV$^4$/$c^8$)}\\
      \rule[-1.3ex]{0pt}{1.3ex}$E^*_\mathrm{min}$ (GeV) & 0.7 & 0.9 &
      1.1 & 1.3 & 1.5 & 1.7 & 1.9\\
      \hline
      \rule{0pt}{2.7ex}secondary/fake leptons & 0.167 & 0.109 & 0.050
      & 0.023 & 0.009 & 0.005 & 0.002\\
      combinatorial background & 0.028 & 0.018 & 0.009 & 0.005 & 0.003
      & 0.002 & 0.001\\
      continuum & 0.000 & 0.000 & 0.000 & 0.000 & 0.000 & 0.001 &
      0.000\\
      \rule[-1.3ex]{0pt}{1.3ex}$B\to X_u\ell\nu$ background & 0.004 &
      0.004 & 0.004 & 0.003 & 0.003 & 0.005 & 0.005\\
      \hline
      \rule{0pt}{2.7ex}$\mathcal{B}(D^{(*)}\ell\nu)$ & 0.013 & 0.010 &
      0.007 & 0.004 & 0.002 & 0.002 & 0.003\\
      $\mathcal{B}(D^{**}\ell\nu)$ & 0.216 & 0.169 & 0.102 & 0.049 &
      0.042 & 0.011 & 0.009\\
      $\mathcal{B}((D^{(*)}\pi)_\mathrm{non-res.}\ell\nu)$ & 0.168 &
      0.125 & 0.058 & 0.041 & 0.024 & 0.004 & 0.004\\
      $D^{(*)}\ell\nu$ form factors & 0.029 & 0.028 & 0.024 & 0.019 &
      0.017 & 0.016 & 0.007\\
      \rule[-1.3ex]{0pt}{1.3ex}$D^{**}\ell\nu$ form factors & 0.013 &
      0.009 & 0.006 & 0.003 & 0.004 & 0.001 & 0.004\\
      \hline
      \rule{0pt}{2.7ex}unfolding & 0.035 & 0.035 & 0.035 & 0.035 &
      0.035 & 0.035 & 0.035\\
      binning & 0.001 & 0.001 & 0.000 & 0.001 & 0.001 & 0.001 &
      0.002\\
      \rule[-1.3ex]{0pt}{1.3ex}efficiency & 0.025 & 0.032 & 0.027 &
      0.014 & 0.013 & 0.005 & 0.002\\
      \hline
      \rule[-1.3ex]{0pt}{4ex}total & 0.327 & 0.244 & 0.137 & 0.080 &
      0.064 & 0.040 & 0.038\\
      \hline \hline
    \end{tabular}
  \end{center}
\end{table}
\begin{table}
  \caption{Same as Table~\ref{tab:3} for $\langle M^4_X\rangle$.}
  \label{tab:5}
  \begin{center}
    \begin{tabular}{l@{\extracolsep{.3cm}}ccccccc}
      \hline \hline
      \rule{0pt}{2.7ex} & \multicolumn{7}{c}{$\Delta\langle
	M^4_X\rangle$ (GeV$^4$/$c^8$)}\\
      \rule[-1.3ex]{0pt}{1.3ex}$E^*_\mathrm{min}$ (GeV) & 0.7 & 0.9 &
      1.1 & 1.3 & 1.5 & 1.7 & 1.9\\
      \hline
      \rule{0pt}{2.7ex}secondary/fake leptons & 0.46 & 0.31 & 0.16 &
      0.09 & 0.04 & 0.02 & 0.00\\
      combinatorial background & 0.08 & 0.05 & 0.04 & 0.03 & 0.01 &
      0.01 & 0.00\\
      continuum & 0.00 & 0.00 & 0.00 & 0.00 & 0.00 & 0.00 & 0.00\\
      \rule[-1.3ex]{0pt}{1.3ex}$B\to X_u\ell\nu$ background & 0.04 & 0.04
      & 0.04 & 0.04 & 0.05 & 0.06 & 0.07\\
      \hline
      \rule{0pt}{2.7ex}$\mathcal{B}(D^{(*)}\ell\nu)$ & 0.07 & 0.07 &
      0.06 & 0.05 & 0.04 & 0.03 & 0.02\\
      $\mathcal{B}(D^{**}\ell\nu)$ & 0.41 & 0.30 & 0.14 & 0.04 & 0.04
      & 0.05 & 0.06\\
      $\mathcal{B}((D^{(*)}\pi)_\mathrm{non-res.}\ell\nu)$ & 0.38 &
      0.28 & 0.12 & 0.08 & 0.06 & 0.04 & 0.04\\
      $D^{(*)}\ell\nu$ form factors & 0.15 & 0.14 & 0.13 & 0.12 & 0.10
      & 0.09 & 0.04\\
      \rule[-1.3ex]{0pt}{1.3ex}$D^{**}\ell\nu$ form factors & 0.04 &
      0.03 & 0.02 & 0.01 & 0.01 & 0.00 & 0.03\\
      \hline
      \rule{0pt}{2.7ex}unfolding & 0.16 & 0.16 & 0.16 & 0.16 & 0.16 &
      0.16 & 0.16\\
      binning & 0.01 & 0.01 & 0.01 & 0.00 & 0.00 & 0.01 & 0.00\\
      \rule[-1.3ex]{0pt}{1.3ex}efficiency & 0.09 & 0.13 & 0.13 & 0.09
      & 0.08 & 0.04 & 0.03\\
      \hline
      \rule[-1.3ex]{0pt}{4ex}total & 0.77 & 0.58 & 0.36 & 0.26 & 0.23
      & 0.21 & 0.20\\
      \hline \hline
    \end{tabular}
  \end{center}
\end{table}

The uncertainties related to the different background components in
$M^2_X$ are estimated by varying the respective background
normalization factors within $\pm 1$~standard deviation.

We consider both variations of the $B\to D^{(*)}\ell\nu$~branching
fractions and form factor shapes. For the former, the ranges of
variation are taken from Ref.~\cite{Yao:2006px}. For the latter,
the curvature~$\rho^2$ in the form factor
parametrization~\cite{Caprini:1997mu} is varied within $1.56\pm 0.14$
($1.15\pm 0.16$) for $B\to D^*\ell\nu$ ($B\to
D\ell\nu$)~\cite{Barberio:2006bi}. For $B\to D^*\ell\nu$, we also vary
the form factor ratios $R_1$ and $R_2$~\cite{Aubert:2006cx}.

The LLSW~model~\cite{Leibovich:1997em} predicts the relative abundance
and the form factor shape of the different components in $B\to
D^{**}\ell\nu$ only. To obtain the absolute branching fractions of the
$B\to D^{**}\ell\nu$~components and of $B\to
(D^{(*)}\pi)_\mathrm{non-res.}\ell\nu$, we use $\mathcal{B}(B^+\to\bar
D_1^0\ell^+\nu)=(5.6\pm 1.6)\times 10^{-3}$~\cite{Yao:2006px},
the recent Belle measurement of $\mathcal{B}(B\to
D^{(*)}\pi\ell\nu)$~\cite{Liventsev:2005hh} and the total semileptonic
branching fraction~\cite{Yao:2006px}. The uncertainty assigned to
the $B\to D^{**}\ell\nu$~branching fractions in
Tables~\ref{tab:3}--\ref{tab:5} reflects the uncertainty in these
measurements and the change in the $B\to D^{**}\ell\nu$~composition
when varying the LLSW parameters within their allowed range.

The SVD~algorithm used to unfold the measured $M^2_X$~distribution
requires the detector response matrix, {\it i.e.}, the distribution of
measured versus true values of $M^2_X$. We determine this matrix from
the MC simulation. To study the systematics related to unfolding and
a possible mismodeling of the detector response, we change the amount
of bin-to-bin migration by varying the effective rank of the detector
response matrix, the main tunable parameter of the SVD algorithm. We
have further studied a change of the binning of the unfolded
distribution and the effect of disabling the bin-to-bin efficiency
correction.

\subsection{Correlations}

Due to overlapping events, the moment measurements corresponding to
different lepton energy thresholds are highly
correlated. Systematic uncertainties are another source of
correlation. We have estimated the correlations due to both sources
using a toy MC approach based on 50,000~simulated measurements. The
results for the self- and cross-correlation coefficients are given in
Tables~\ref{tab:6}--\ref{tab:10}.
\begin{table}
  \caption{Correlation coefficients between $\langle
    M^2_X\rangle$~measurements.} \label{tab:6}
  \begin{center}
    \begin{tabular}{c@{\extracolsep{.3cm}}c|ccccccc}
      \hline \hline
      \multicolumn{2}{c|}{\rule{0pt}{2.7ex}$E^*_\mathrm{min}$} &
      \multicolumn{7}{c}{$\langle M^2_X\rangle$}\\
      \multicolumn{2}{c|}{\rule[-1.3ex]{0pt}{1.3ex}(GeV)} & 0.7 & 0.9 &
      1.1 & 1.3 & 1.5 & 1.7 & 1.9\\
      \hline
      \rule{0pt}{2.7ex} & 0.7 & 1.000 & 0.932 & 0.786 & 0.615 & 0.481
      & 0.168 & 0.071\\
      & 0.9 & & 1.000 & 0.888 & 0.715 & 0.573 & 0.241 & 0.116\\
      & 1.1 & & & 1.000 & 0.849 & 0.693 & 0.363 & 0.194\\
      $\langle M^2_X\rangle$ & 1.3 & & & & 1.000 & 0.804 & 0.470 &
      0.254\\
      & 1.5 & & & & & 1.000 & 0.591 & 0.308\\
      & 1.7 & & & & & & 1.000 & 0.363\\
      \rule[-1.3ex]{0pt}{1.3ex} & 1.9 & & & & & & & 1.000\\
      \hline \hline
    \end{tabular}
  \end{center}
\end{table}
\begin{table}
  \caption{Correlation coefficients between $\langle
    M^2_X\rangle$ and $\langle (M^2_X-\langle
    M^2_X\rangle)^2\rangle$~measurements.}
  \begin{center}
    \begin{tabular}{c@{\extracolsep{.3cm}}c|ccccccc}
      \hline \hline
      \multicolumn{2}{c|}{\rule{0pt}{2.7ex}$E^*_\mathrm{min}$} &
      \multicolumn{7}{c}{$\langle (M^2_X-\langle
      M^2_X\rangle)^2\rangle$}\\
      \multicolumn{2}{c|}{\rule[-1.3ex]{0pt}{1.3ex}(GeV)} & 0.7 & 0.9 &
      1.1 & 1.3 & 1.5 & 1.7 & 1.9\\
      \hline
      \rule{0pt}{2.7ex} & 0.7 & 0.897 & 0.847 & 0.788 & 0.713 & 0.576
      & 0.306 & 0.102\\
      & 0.9 & 0.777 & 0.843 & 0.804 & 0.726 & 0.608 & 0.356 & 0.144\\
      & 1.1 & 0.548 & 0.615 & 0.757 & 0.690 & 0.606 & 0.426 & 0.211\\
      $\langle M^2_X\rangle$ & 1.3 & 0.328 & 0.371 & 0.483 & 0.718 &
      0.599 & 0.476 & 0.260\\
      & 1.5 & 0.223 & 0.263 & 0.346 & 0.481 & 0.702 & 0.559 & 0.280\\
      & 1.7 & $-.051$ & $-.031$ & 0.035 & 0.126 & 0.237 & 0.846 &
      0.296\\
      \rule[-1.3ex]{0pt}{1.3ex} & 1.9 & $-.060$ & $-.047$ & $-.007$ &
      0.040 & 0.075 & 0.228 & 0.865\\
      \hline \hline
    \end{tabular}
  \end{center}
\end{table}
\begin{table}
  \caption{Correlation coefficients between $\langle
    M^2_X\rangle$ and $\langle M^4_X\rangle$~measurements.}
  \begin{center}
    \begin{tabular}{c@{\extracolsep{.3cm}}c|ccccccc}
      \hline \hline
      \multicolumn{2}{c|}{\rule{0pt}{2.7ex}$E^*_\mathrm{min}$} &
      \multicolumn{7}{c}{$\langle M^4_X\rangle$}\\
      \multicolumn{2}{c|}{\rule[-1.3ex]{0pt}{1.3ex}(GeV)} & 0.7 & 0.9 &
      1.1 & 1.3 & 1.5 & 1.7 & 1.9\\
      \hline
      \rule{0pt}{2.7ex} & 0.7 & 0.983 & 0.933 & 0.830 & 0.683 & 0.523
      & 0.194 & 0.073\\
      & 0.9 & 0.890 & 0.974 & 0.910 & 0.765 & 0.606 & 0.264 & 0.117\\
      & 1.1 & 0.704 & 0.810 & 0.976 & 0.857 & 0.707 & 0.380 & 0.196\\
      $\langle M^2_X\rangle$ & 1.3 & 0.508 & 0.601 & 0.774 & 0.980 &
      0.800 & 0.479 & 0.258\\
      & 1.5 & 0.383 & 0.469 & 0.614 & 0.764 & 0.985 & 0.597 & 0.305\\
      & 1.7 & 0.079 & 0.137 & 0.273 & 0.403 & 0.539 & 0.994 & 0.357\\
      \rule[-1.3ex]{0pt}{1.3ex} & 1.9 & 0.017 & 0.052 & 0.136 & 0.208
      & 0.271 & 0.348 & 0.995\\
      \hline \hline
    \end{tabular}
  \end{center}
\end{table}
\begin{table}
  \caption{Correlation coefficients between $\langle (M^2_X-\langle
    M^2_X\rangle)^2\rangle$~measurements.}
  \begin{center}
    \begin{tabular}{c@{\extracolsep{.3cm}}c|ccccccc}
      \hline \hline
      \multicolumn{2}{c|}{\rule{0pt}{2.7ex}$E^*_\mathrm{min}$} &
      \multicolumn{7}{c}{$\langle (M^2_X-\langle
	M^2_X\rangle)^2\rangle$}\\
      \multicolumn{2}{c|}{\rule[-1.3ex]{0pt}{1.3ex}(GeV)} & 0.7 & 0.9 &
      1.1 & 1.3 & 1.5 & 1.7 & 1.9\\
      \hline
      \rule{0pt}{2.7ex} & 0.7 & 1.000 & 0.939 & 0.838 & 0.698 & 0.534
      & 0.167 & $-.024$\\
      & 0.9 & & 1.000 & 0.901 & 0.732 & 0.586 & 0.195 & $-.011$\\
      $\langle(M^2_X-$ & 1.1 & & & 1.000 & 0.793 & 0.638 & 0.262 &
      0.034\\
      $-\langle M^2_X\rangle)^2\rangle$ & 1.3 & & & & 1.000 & 0.731 &
      0.340 & 0.102\\
      & 1.5 & & & & & 1.000 & 0.484 & 0.146\\
      & 1.7 & & & & & & 1.000 & 0.296\\
      \rule[-1.3ex]{0pt}{1.3ex} & 1.9 & & & & & & & 1.000\\
      \hline \hline
    \end{tabular}
  \end{center}
\end{table}
\begin{table}
  \caption{Correlation coefficients between $\langle
    M^4_X\rangle$~measurements.}
  \label{tab:10}
  \begin{center}
    \begin{tabular}{c@{\extracolsep{.3cm}}c|ccccccc}
      \hline \hline
      \multicolumn{2}{c|}{\rule{0pt}{2.7ex}$E^*_\mathrm{min}$} &
      \multicolumn{7}{c}{$\langle M^4_X\rangle$}\\
      \multicolumn{2}{c|}{\rule[-1.3ex]{0pt}{1.3ex}(GeV)} & 0.7 & 0.9 &
      1.1 & 1.3 & 1.5 & 1.7 & 1.9\\
      \hline
      \rule{0pt}{2.7ex} & 0.7 & 1.000 & 0.932 & 0.784 & 0.601 & 0.442
      & 0.111 & 0.017\\
      & 0.9 & & 1.000 & 0.877 & 0.684 & 0.524 & 0.168 & 0.051\\
      & 1.1 & & & 1.000 & 0.817 & 0.651 & 0.297 & 0.137\\
      $\langle M^4_X\rangle$ & 1.3 & & & & 1.000 & 0.780 & 0.421 &
      0.212\\
      & 1.5 & & & & & 1.000 & 0.557 & 0.270\\
      & 1.7 & & & & & & 1.000 & 0.346\\
      \rule[-1.3ex]{0pt}{1.3ex} & 1.9 & & & & & & & 1.000\\
      \hline \hline
    \end{tabular}
  \end{center}
\end{table}

\section{Summary}

We have measured the first, $\langle M^2_X\rangle$, and the second
central and non-central moments, $\langle (M^2_X-\langle
M^2_X\rangle)^2\rangle$ and $\langle M^4_X\rangle$, of the hadronic
mass squared spectrum in $B\to X_c\ell\nu$~decays for lepton energy
thresholds ranging from 0.7 to 1.9~GeV. Using a toy MC approach, we
have also evaluated the full covariance matrix for this set of
measurements.

It is expected that this measurement, combined with measurements of
the semileptonic branching fraction, moments of the lepton energy
spectrum in $B\to X_c\ell\nu$~decays and possibly other moments, will
lead to an improved determination of $b$-quark mass~$m_b$ and the CKM
matrix
element~$|V_{cb}|$~\cite{Benson:2003kp,Gambino:2004qm,Benson:2004sg,Bauer:2002sh}.

\section*{Acknowledgments}
% Please paste this acknowledgement into your latex file. 
% updated 11/22/06  Chinese Academy of Sciencies -> Sciences
% updated 6/23/06
% updated 2/21/06
% updated 1/28/06
% updated 12/25/05
% short version reduced 8/11/05
% updated 7/17/05
% updated 2/17/05
%***** Acknowledgments *****
%----------- Long version, for most papers ----------- 
We thank the KEKB group for the excellent operation of the
accelerator, the KEK cryogenics group for the efficient
operation of the solenoid, and the KEK computer group and
the National Institute of Informatics for valuable computing
and Super-SINET network support. We acknowledge support from
the Ministry of Education, Culture, Sports, Science, and
Technology of Japan and the Japan Society for the Promotion
of Science; the Australian Research Council and the
Australian Department of Education, Science and Training;
the National Science Foundation of China and the Knowledge
Innovation Program of the Chinese Academy of Sciences under
contract No.~10575109 and IHEP-U-503; the Department of
Science and Technology of India; 
the BK21 program of the Ministry of Education of Korea, 
the CHEP SRC program and Basic Research program 
(grant No.~R01-2005-000-10089-0) of the Korea Science and
Engineering Foundation, the Pure Basic Research Group 
program of the Korea Research Foundation, and the SBS Foundation;
the Polish State Committee for Scientific Research; 
%-> remove for now: under contract No.~2P03B 01324; 
the Ministry of Science and Technology of the Russian
Federation; the Slovenian Research Agency;  the Swiss
National Science Foundation; the National Science Council
and the Ministry of Education of Taiwan; and the U.S.\
Department of Energy.


\begin{thebibliography}{99}
\bibitem{ref:2}
  In this paper, the inclusion of the charge conjugate decay is
  implied.
\bibitem{ref:3}
  In this paper, quantities calculated in the $B$~meson rest frame are
  denoted by an asterisk.
\bibitem{ref:4}
  These numbers are our private averages of $B\to
  X_u\ell\nu$~measurements by LEP, CLEO, BaBar and Belle. The
  uncertainty in these branching fractions is assumed to be 20\%.
\end{thebibliography}
\end{document}